\documentclass[twocolumn,showpacs,preprintnumbers,amsmath,amssymb,prb,aps]{revtex4-1}
\usepackage{epsfig}
\usepackage{epstopdf}
\usepackage{graphicx}
\usepackage{dcolumn}
\usepackage{bm}
\usepackage{textcomp}
\usepackage{array}
\usepackage{subcaption}
\usepackage{booktabs}
\usepackage{natbib}
\usepackage{url}
\begin{document}

\title{Thermoelectric properties, efficiency and thermal expansion of ZrNiSn half-Heusler by first-principles calculations }
\author{Shivprasad S. Shastri}
\altaffiliation{Electronic mail: shastri1992@gmail.com}
\author{Sudhir K. Pandey}
\altaffiliation{Electronic mail: sudhir@iitmandi.ac.in}
\affiliation{School of Engineering, Indian Institute of
Technology Mandi, Kamand - 175005, India}

\date{\today}

\begin{abstract}

 In this work, we try to understand the experimental thermoelectric (TE) properties of a ZrNiSn sample with DFT and semiclassical transport calculations using SCAN functional. SCAN and mBJ provide the same band gap $E_{g}$ of $\sim$0.54 eV. This $E_{g}$ is found to be inadequate to explain the experimental data. The better explanation of experimental Seebeck coefficient $S$ is done by considering $E_{g}$ of 0.18 eV which suggests the non-stoichiometry and/or disorder in the sample. Further improvement in the $S$ is done by the inclusion of temperature dependence on chemical potential. In order to look for the possible enhanced TE properties obtainable in ZrNiSn with $E_{g}$ of $\sim$0.54 eV, power factor and optimal carrier concentrations are calculated. The optimal electron and hole concentrations required to attain highest power factors are $\sim$7.6x10$^{19}$ cm$^{-3}$ and $\sim$1.5x10$^{21}$ cm$^{-3}$, respectively. The maximum figure of merit $ZT$ calculated at 1200 K for n-type and p-type ZrNiSn are $\sim$0.6 and $\sim$0.7,  respectively. The \% efficiency obtained for n-type ZrNiSn is $\sim$5.1 \% while for p-type ZrNiSn is $\sim$6.1 \%. The $ZT$ are expected to be further enhanced to $\sim$1.2 (n-type) and $\sim$1.4 (p-type) at 1200 K by doping with heavy elements for thermal conductivity reduction. The phonon properties are also studied by calculating dispersion, total and partial density of states. The calculated Debye temperature of 382 K is in good agreement with experimental value of 398 K. The thermal expansion behaviour in ZrNiSn is studied under quasi-harmonic approximation. The average linear thermal expansion coefficient $\alpha_{ave}(T)$ of $\sim$7.8x10$^{-6}$ K$^{-1}$ calculated in our work is quite close to the experimental values. The calculated linear thermal expansion coefficient will be useful in designing the thermoelectric generators for high temperature applications.

\end{abstract}

\maketitle

\section{Introduction} 
Thermoelectric materials are one of the technologically important materials for small scale power generation applications through waste heat recovery. In a thermoelectric material generation of electricity takes place by the motion of charge carriers driven by flow of heat across a temperature gradient.\cite{small,compintro} This  process in the solid state thermoelectric device leads to power generation useful enough for many applications. At many places where heat released is unused, like in automobiles, industry, many heating appliances in homes or offices, high performance computing (HPC) centers etc. thermoelectric generators (TEG) can be applied. This helps in providing an alternative to conventional fuel sources, usefulness of waste heat through electricity generation and lesser impact on environment. The advantages of thermoelectric materials at applications are that they are scalable according to application, portable, lesser in weight, noise-free and less expensive.\cite{tritt} 

The dimensionless figure of merit $ZT$ for a thermoelectric material is given by the relation, 
 \begin{equation}
ZT = \frac{S^{2}\sigma T}{\kappa_{e}+\kappa_{ph}}.
\end{equation} 
The above relation gives a measure of efficiency of a thermoelectric material. As the equation suggests, to have high a $ZT$ the Seebeck coefficient $S$, electrical conductivity $\sigma$ should be high and electronic part and phonon part of thermal conductivity ($\kappa_{e}$ and $\kappa_{ph}$, respectively) should be lower at any temperature $T$. But, the quantities deciding the $ZT$ are counter related and hence achieving good value of $ZT$ is a task of optimization among these quantities.\cite{snydercomplex,mahanbest} Also, there is a constant effort in the areas of thermoelectric research to reduce the thermal conductivity by methods like nanostructuring, alloying and to enhance $S$ and $\sigma$ by suitable doping.\cite{snydercomplex} With the help of density functional theory (DFT) and computational tools, materials with suitable electronic structure are being discovered which can be further tuned to get good thermoelectric efficiency.\cite{yangintro} Materials with $ZT$ value closer to or greater than 1.0 can be considered for application in commercial TEGs.

The ability of a TEG to convert heat into electrical power is given by its efficiency $\eta$. The maximum efficiency $\eta_{max}$ of a TEG is given by\cite{maxeff,gaurav} 
\begin{equation}
\eta_{max} = \frac{T_{h}-T_{c}}{T_{h}} \frac{\sqrt{1+Z\overline{T}}-1}{\sqrt{1+Z\overline{T}}+ T_{c}/T_{h}}.
\end{equation}
Here, $T_{h}$ is hot end temperature, $T_{c}$ is cold end temperature, $Z\overline{T}$ is the figure of merit of the thermoelectric material used in the TEG and $\overline{T}$ is average temperature. The above formula suggests that efficiency of TEG is mainly dependent on temperature difference between hot and cold end and also on the $ZT$ of the material used in the TEG.\cite{gaurav,small} The TEG consists of number of n-type and p-type thermoelements (or legs) connected electrically in series. In order to have good efficiency, materials with high $ZT$ are desirable to use as legs of TEG. The efficiency of TEGs are generally in the range $\sim$7-9 \% depending on the temperature range.\cite{gaurav} Even though the efficiency of TEG is not quite high still they are suitable in applications where sources of energy (\textit{e.g} space applications RTEGs), reliability and silent operation are more important.\cite{dressel} For a practical application of a material in TEG, apart from its high value of $ZT$, considering the properties such as bulk modulus, thermal expansion and melting point etc. are also important. The material with high melting point, hardness and high bulk modulus, and less expensive are advantageous from the application point of view.

 The possession of narrow bands with high Seebeck coefficient, tunable band gap and charge carrier concentration are the features making the Heusler compounds to be explored for thermoelectric applications with the only hindering factor being their relatively high thermal conductivity.\cite{felser} ZrNiSn belongs to the family of half-Heusler compounds with valence electron count 18 and has semiconducting ground state.\citep{rabe} The experimentally reported band gap for ZrNiSn by the resistivity measurement is $\sim$0.18 eV.\cite{bandgap} The ZrNiSn compound shows high $S$ values and moderately high value of $\sigma$ which make it promising material to explore for high temperature thermoelectric application. But, the $\kappa$ of the undoped ZrNiSn is $\sim$10-6 Wm$^{-1}$K$^{-1}$ (300-900 K) which is relatively higher compared to oxide or telluride materials.\cite{shenexp,snydertrue} Due to the higher power factor in ZrNiSn based compounds, much attention has been given in exploring the possibility to enhance the $ZT$. Shen \textit{et al.}\cite{shenexp} reported the thermoelectric properties of undoped and Pd substituted ZrNiSn compounds. A maximum $ZT$ value of $\sim$0.7 at 800 K was observed for the 
Hf$_{0.5}$Zr$_{0.5}$Ni$_{0.8}$Pd$_{0.2}$Sn$_{0.99}$Sb$_{0.01}$ compound. Sb doping at the Sn site in ZrNiSn was found to enhance the $ZT$ value to $\sim$0.8 at 800 K in the work of Culp \textit{et al.}\cite{culp} In the n-type (Hf,Zr)NiSn phase samples Chen \textit{et al.} have observed a $ZT$ of $\sim$1.2 at $\sim$800 K through proper annealing.\cite{chen} Theoretically, Zou \textit{et al.} have given the explanation of experimental thermoelectric properties of doped ZrNiSn and Zr$_{0.5}$Hf$_{0.5}$NiSn compound using DFT and transport calculations.\cite{zouelectronic} But, an attempt to understand the experimental thermoelectric properties of undoped ZrNiSn samples is lacking in the literature. Moreover, high temperature thermal expansion behaviour becomes an important parameter when designing TEG which is calculated here and compared with experimental data.   

Therefore, in the present work, the experimental thermoelectric properties of ZrNiSn compound\cite{shenexp} is explained using DFT and transport calculations. SCAN functional is used to study the electronic structure and transport properties. To explain the thermoelectric properties, temperature dependent shift in chemical potential is considered. Further, the $S$, $\sigma$ and $\kappa_{e}$ of ZrNiSn with band gap 0.54 eV are predicted. Importance of considering doping in the ordered ZrNiSn with band gap of 0.54 eV is shown by exploring the possibility of improving $ZT$ upto $\sim$1.4 for p-type doping.  We have calculated the optimal doping concentration which would yield highest power factor for n-type and p-type material which are $\sim$7.6x10$^{19}$ cm$^{-3}$ and $\sim$1.5x10$^{21}$ cm$^{-3}$, respectively. The figure of merit for n-type and p-type ZrNiS are estimated which are $\sim$0.6 and $\sim$0.7 at 1200 K, respectively. Applying the method of segmentation, efficiency of n-type and p-type materials are estimated. The obtained efficiency of p-type material is $\sim$6.1 \% and that of n-type material is $\sim$5.1 \% for cold and hot end temperature of 300 K and 1200K, respectively. By doping with heavy elements to reduce $\kappa$ for the ordered ZrNISn the $ZT$ are predicted to be enhanced to $\sim$1.4 and $\sim$1.2 for p-type and n-type compounds, respectively at 1200 K. Thus, through this work, emphasis of preparing doped ZrNiSn samples without disorder is theoretically studied by exploring the possible enhancement in $ZT$. The phonon properties are studied by calculating phonon dispersion, total and partial density of states. The estimated Debye temperature is $\sim$382 K which is close to the experimental value of 398 K. Under the qhasi-harmonic approximation, the thermal expansion of ZrNiSn is studied. The calculated average linear thermal expansion coefficient $\alpha_{ave}(T)$ is $\sim$7.8x10$^{-6}$ K$^{-1}$. This value is in quite good agreement with the experimentally reported value  of 11.0x10$^{-6}$ K$^{-1}$ from the high temperature XRD measurement.

\section{Computational details}
The DFT calculations are performed using full-potential linearized augmented plane wave (FP-LAPW) method based WIEN2k code.\cite{wien2k} Using WIEN2k, the ground state total energy, electronic structure and forces on atoms are calculated. Meta-GGA SCAN\cite{scan} is used as the exchange-correlation (XC) functional in this work. For the convergence of electronic total energy, a criterion of 10$^{-4}$ Ry/cell is selected. In the ground-state total energy calculations, a dense k-point mesh of 50x50x50 is used in order to get computationally proper transport properties further. The transport properties of the compound are calculated using BoltzTraP\cite{boltztrap} program based on semiclassical Boltzmann transport theory. The doping in BoltzTraP is considered through the shift in chemical potential. The lattice parameter used in this work is optimized by fitting Birch-Murnaghan (BM) EOS\cite{birch} to energy vs. volume curves.

The force constants and phonon properties are calculated using phonopy\cite{togo} under supercell and finite displacement method (FDM). In order to calculate the forces on atoms in WIEN2k, a supercell of size 2x2x2 is constructed of the conventional unit cell.  Under an approximation termed quasi-harmonic approximation (QHA) as implemented in phonopy\cite{togo} we calculate thermal expansion of ZrNiSn. In QHA volume dependence on the phonon frequencies are introduced through harmonic approximation in order to get the thermal properties. 

Electronic total energy and force calculations are carried out at 21 volumes points varying from -5\% to +5\% of equilibrium unit cell volume. In order to find out volume corresponding to minimum total free energy at different temperatures (0K to 1300 K in steps of 100 K), BM EOS is used at each temperature (Fig. 9 (a)). The Brillouin zones of the supercells are sampled  by 5x5x5 k-point mesh. A force convergence criteria of 0.1mRy/bohr is used in the force calculator. 

\begin{figure}
\includegraphics[width=6cm, height=8cm]{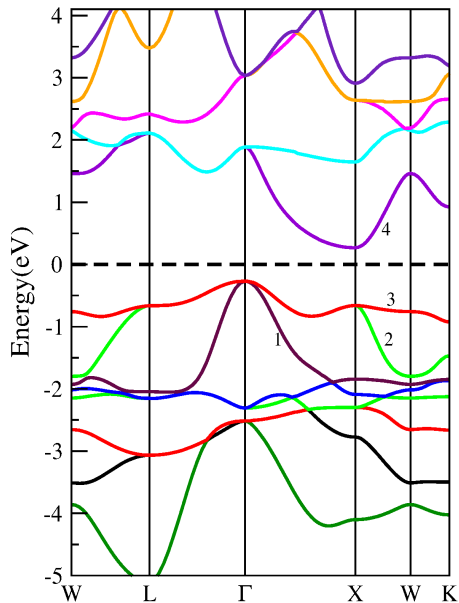} 
\caption{The electronic dispersion of ZrNiSn compound.}
\end{figure}

\section{Results and Discussion}
The ZrNiSn compound belongs to the class of half-Heuslers and has cubic $C1_{b}$ structure with space group $F\overline{4}3m$ (No. 216). In half-Heusler structure theoretically three inequivalent atomic arrangements are possible.\citep{felser} The calculated total energy of other two possible atomic arrangements are $\sim$2.5 eV/f.u and $\sim$3.0 eV/f.u higher, respectively with respect to ground state of the compound which is semi-conducting. In the ground state structure the Wyckoff positions of Zr, Ni and Sn atoms are $4b$ (1/2, 1/2, 1/2), $4c$ (1/4, 1/4, 1/4) and $4a$ (0, 0, 0), respectively. The ground state structure is used for the further electronic structure, transport properties and phonon calculations. The experimental value of lattice constant taken from the literature\cite{latticeconstant2,latticeconstant1} is 6.110 \AA. The volume optimization procedure is carried out to find out the ground state equilibrium  lattice constant. The equilibrium lattice constant after fitting the BM equation of state to the total energy $vs.$ unit cell volume curve is 6.092 \AA. The calculated lattice constant using SCAN XC functional is in close agreement with the experimental value. 
\begin{figure*}
\includegraphics[width=16cm, height=12cm]{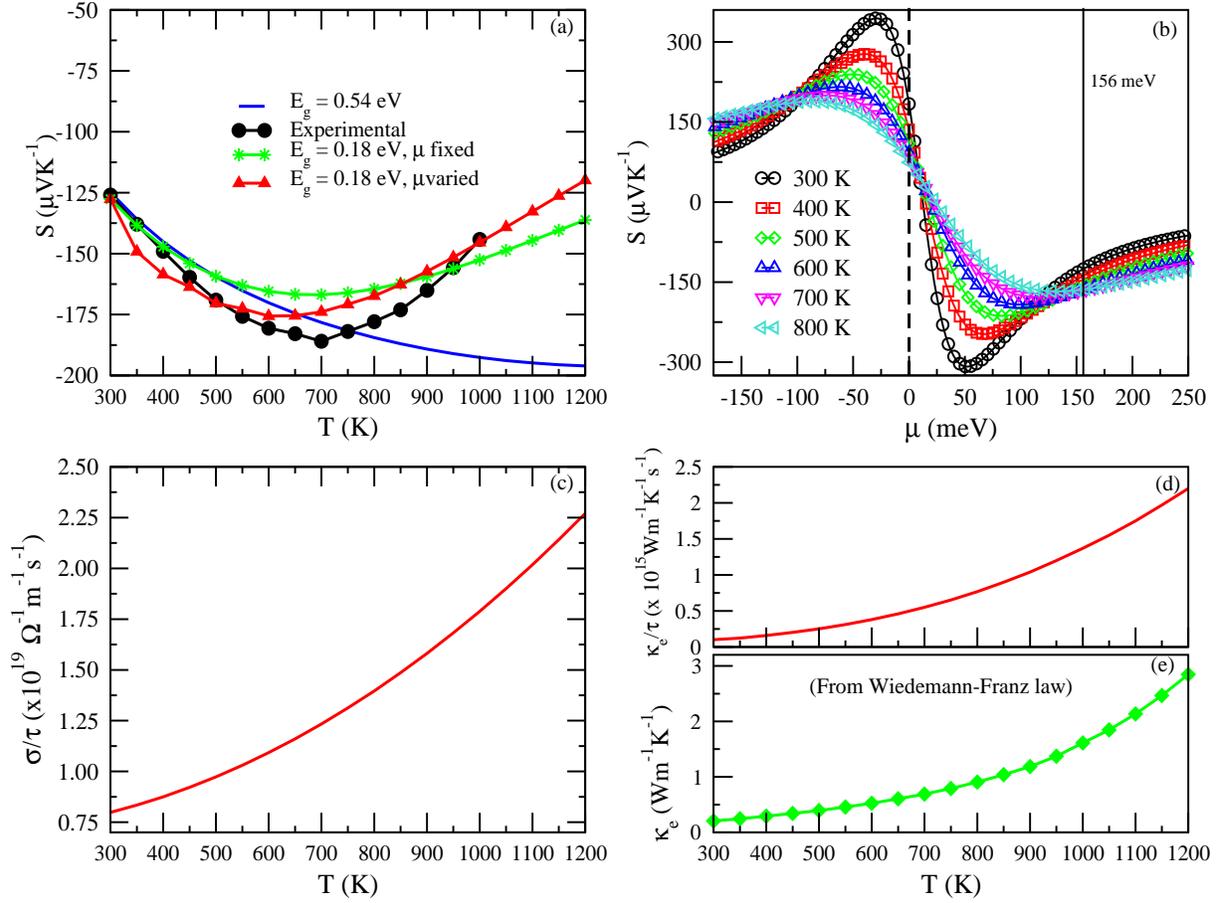} 
\caption{ (a) The experimental and calculated Seebeck coefficient $S$ with temperature. The experimental $S$ is from Ref. 15. (b) Seebeck coefficient ($S$) for change in chemical potential $\mu$ for band gap of 0.18 eV.  (c) Electrical conductivity per relaxation time ($\sigma$/$\tau$) (d) electronic thermal conductivity per relaxation time ($\kappa_{e}$/$\tau$) with temperature and (e) electronic thermal conductivity $\kappa_{e}$ extracted from experimental data.}
\end{figure*}

\subsection{\label{sec:level2}Electronic structure}
The electronic dispersion of ZrNiSn is calculated using SCAN XC functional which is presented in Fig. 1. The SCAN is found to describe the electronic dispersion features (group velocity, effective mass) more appropriately relative to LDA, PBE or mBJ and thus giving improved transport properties.\cite{shamim,paper3} As can be seen from the electronic dispersion the ZrNiSn is a semiconductor with an indirect band gap of 0.54 $eV$. For this compound, mBJ\cite{mbj} calculation performed is also giving the value of indirect band gap of 0.55 $eV$. The values of band gaps obtained from SCAN and mBJ match very well with each other and also to the earlier reported value from LDA and PBE.\cite{zouelectronic,rabe} In the figure dashed line in the middle of the gap represents the Fermi level ($E_{F}$) and the bands that would contribute significantly to transport are numbered from 1 to 4 (referred with symbols B1 to B4 from here on and in Table I). In the ZrNiSn compound, the conduction band minimum (CBM) is at the $X$-point and the triply degenerate valence band maximum (VBM) at the $\Gamma$ point which was also seen in case of full-Heusler compounds studied in our earlier work\cite{paper3}. The degeneracy of bands B2 and B3 can be observed fully along the $L$-$\Gamma$-$X$ k-path. This is a useful feature for a thermoelectric material since, more states are available for occupation of charge carriers near the chemical potential. Also, the bands B2 and B3 are narrower causing charge carriers to have more effective mass and thus higher contribution to Seebeck coefficient. The lift in degeneracy of the band B1 with other two bands can be seen on moving away from the vicinity of $\Gamma$-point. 
\begin{figure*}
\includegraphics[width=16cm, height=12cm]{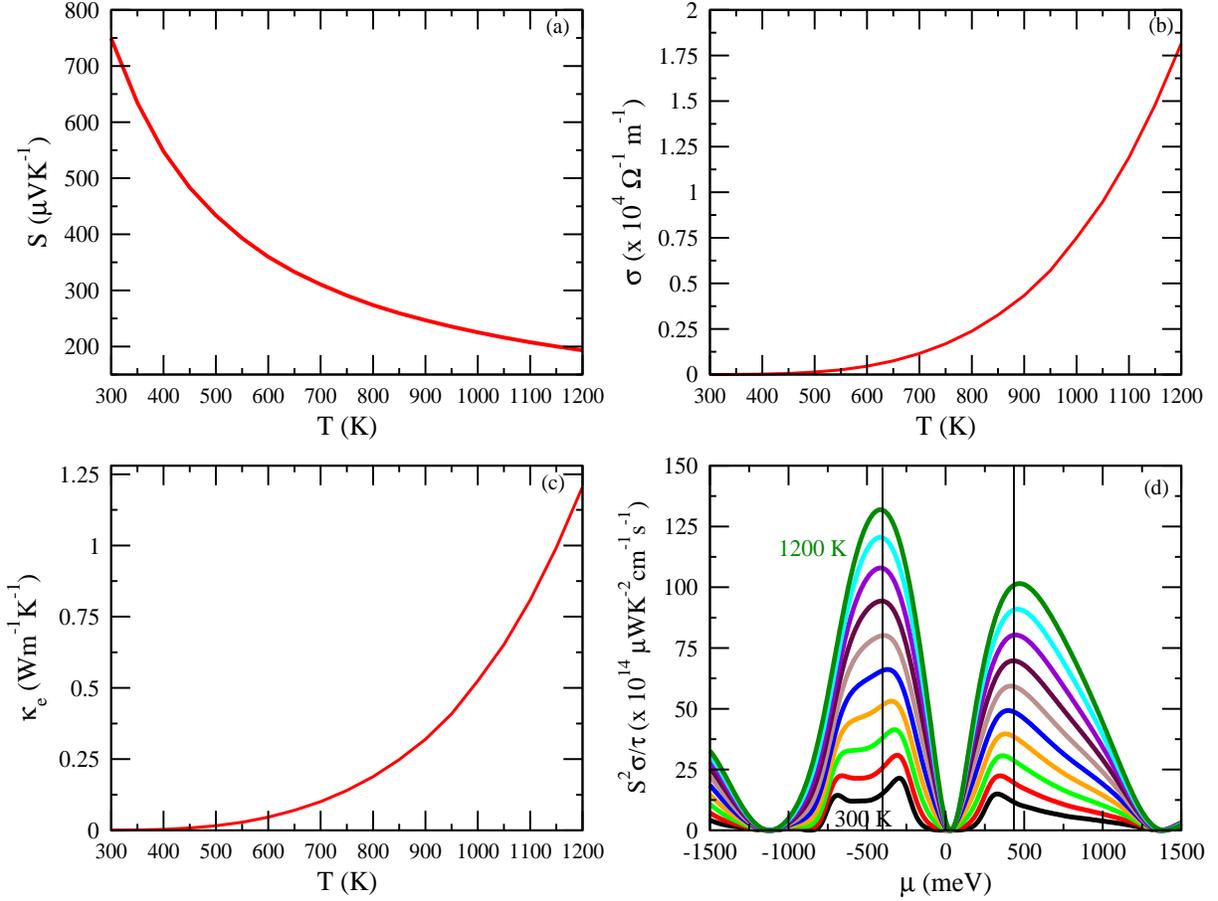} 
\caption{The variation of (a) Seebeck coefficient $S$, (b) electrical conductivity $\sigma$  and (c) electronic thermal conductivity $\kappa_{e}$ with temperature for ZrNiSn (E$_{g}$=0.54 eV). (d) Power factor per relaxation time with change in chemical potential $\mu$.  }
\end{figure*}

The effective mass $m^{*}$ of charge carriers in a material is one of the key quantities for understanding the transport behaviour of that material. The $m^{*}$ values of charge carriers in terms of mass of electron ($m_{e}$) for bands B1 to B4 are calculated and listed in Table I. In order to find the $m^{*}$, parabola is fitted in the neighbourhood of band extrema where the charge carriers behave more like free particles. In the table, the notation for instance, $\Gamma$-$\Gamma$L denotes the $m^{*}$ calculated in the vicinity of $\Gamma$ point along $\Gamma$ to L direction. Clearly, the band B1 is lighter band with lesser $m^{*}$ and the charge carrier conductivity will be higher in this band. The bands B2 and B3 being degenerate have same $m^{*}$ values in all the directions listed except in X-W direction. The value of $m^{*}$ for B3 (and B2 is) are $\sim$2.42 and $\sim$2.50 times larger than $m_{e}$ in $\Gamma$-L and X-W directions, respectively. The effective mass for the conduction band B4 is $\sim$3.56 along X-$\Gamma$ direction indicating lower mobility for electrons in the vicinity of CBM in ZrNiSn. 

\subsection{\label{sec:level2}Transport properties}
In this section, the Seebeck coefficient $S$ and other electrical transport properties of ZrNiSn compound are explained through transport calculations. Using the calculated values of $S$, an understanding of the experimental $S$ of the ZrNiSn compound is made. The experimental data of ZrNiSn compound is taken from the work of Shen \textit{et al.}\cite{shenexp} which is presented Fig. 2 (a). At temperature of 300 K, the calculated value of $S$ is $\sim$574 $\mu$VK$^{-1}$ for the ZrNiSn compound. But, the calculated value is positive and far from the experimental value of $\sim$ -125 $\mu$VK$^{-1}$ at the same temperature.  
In the experimentally prepared Huesler compounds, generally non-stoichiometry or inter-mixing between two sites are observed. This can make the position of the chemical potential $\mu$ to shift from the middle of the gap. Therefore, considering the change in $\mu$ may be required in order to explain the observed negative value of experimental $S$. On taking into account the shift in $\mu$ from the middle of the gap, the best matching corresponding to experimental $S$ value was found at $\mu$ of $\sim$337 meV towards conduction band region. The Seebeck coefficient values corresponding to the $\mu$ level of $\sim$337 meV are plotted in Fig. 2 (a). As can be observed from the figure, the calculated $S$ values are close to experimental values upto $\sim$500 K and above 500 K starts to deviate from the experimental curve. Also, the magnitude of calculated $S$ is showing an increasing trend upto 1200 K whereas the experimental data is showing increasing trend only upto $\sim$700 K and above which reduction in $\vert S\vert$ values are observed.

At this point it is important to note that the theoretically obtained band gap (0.54 eV) is higher compared to experimental band gap\cite{bandgap} of 0.18 eV and $S$ of a material is highly dependent on the value of band gap. This may be the reason for the observed differences in the calculated and experimental $S$. Thus in order to explore this possibility and to understand the observed deviation with experimental $S$, the band gap is set to the experimentally observed band gap $E_{g}$ of 0.18 eV.\cite{bandgap} The reduction in $E_{g}$ is considered by a rigid shift of conduction bands in transport properties calculation. The variation of $S$ with doping level ($\mu$-level) for ZrNiSn with $E_{g}$ of 0.18 eV is plotted in Fig. 2(b). In the figure, the dashed line represents chemical potential, $\mu = 0$ at the middle of the gap for intrinsic ZrNiSn at T = 0 K. The chemical potentials for the n-type and p-type compound with the addition of impurities or doping are represented by positive and negative values in the $\mu$-axis, respectively. From the figure one can observe that the best agreement with the experimental value of $S$ is found for $\mu$  of $\sim$156 meV at 300 K. The calculated values of $S$ for the $\mu$ of $\sim$156 meV are plotted in Fig. 2(a). The nature of the $S$ curve thus calculated is showing improvement qualitatively (non-monotonic temperature dependence) compared to that of $S$ curve calculated with $E_{g}$ of 0.54 eV. But, the values of $S$ are still further away from the experimental value. Here, it is important to note that the  $\mu$ level considered till now in calculating $S$ are only doping level dependent and not temperature dependent (fixed $\mu$).  This suggests that considering fixed $\mu$ at all temperatures may be the reason for the observed deviations. In order to verify this we move on to include the temperature dependence on $\mu$ along with the doping dependence.

It is known that the chemical potential in a material with a finite band gap varies with the addition of impurity levels and/or temperature. In an intrinsic semiconductor the chemical potential $\mu_{i}$ at a given temperature T is given by the relation:
\begin{equation}
\mu_{i} = E_{v} + \frac{1}{2}E_{g} + \frac{3}{4}k_{B}T ln \left(\frac{m_{v}}{m_{c}}\right)
\end{equation}
where, $E_{v}$ is energy of the top of the valence band (VBM), $E_{g}$ is energy band gap, T is absolute temperature, $k_{B}$ is Boltzmann constant, $m_{v}$ and $m_{c}$ are effective mass at the VBM and CBM, respectively.\cite{ashcroft,saurabh}  We employed the above relation to include temperature dependence on $\mu$ to calculate the $S$, electrical conductivity per relaxation time ($\sigma$/$\tau$) and electronic thermal conductivity per relaxation time ($\kappa_{e}$/$\tau$) at $\mu$ of 156 meV to further see the nature of calculated $S$. The values for $m_{v}$ and $m_{c}$ at the $\Gamma$ and X-points are taken from Table I to calculate temperature dependent $\mu$. The Seebeck coefficient thus obtained is shown in Fig. 2 (a). The nature of the $S$ curve with $\mu$ varied has improved with respect to the one calculated with fixed $\mu$. This result suggests that considering temperature dependent $\mu$ in calculations could explain better the $S$ behaviour of a compound although in our case calculated $S$ values by this method are deviating at higher temperatures. The possible reason for this deviation may be temperature dependence of band gap\cite{singhgap} and dispersion curves which are not considered in the rigid band approximation.

In Fig. 2 (c), $\sigma$/$\tau$ calculated by considering change in $\mu$ with temperature is shown. The increasing trend of $\sigma$/$\tau$ in 300-1200 K range can be seen. The $\sigma$/$\tau$ value is changing from $\sim$0.78x10$^{19}$ to $\sim$2.25x10$^{19}$ $\Omega^{-1}m^{-1}s^{-1}$ in the temperature range considered. The trend of calculated $\sigma$/$\tau$ by including temperature dependence on $\mu$ is in good agreement with the trend of experimental conductivity.\cite{shenexp} Also, the $\kappa_{e}$/$\tau$ for ZrNiSn is calculated  by considering the temperature dependence on $\mu$ which is shown in Fig. 2 (d). We also extracted the electronic thermal conductivity $\kappa_{e}$, from the experimental value of resistivity\cite{shenexp} using the Wiedemann-Franz law in order to compare with calculated $\kappa_{e}$/$\tau$ and to further get the $\kappa_{ph}$ part of ZrNiSn from the experimental $\kappa$. The nature of the calculated $\kappa_{e}$/$\tau$ qualitatively agrees with the experimental $\kappa_{e}$ of the sample plotted in Fig 2 (e).

The above discussion of the results suggests that better understanding of the transport coefficient of the ZrNiSn sample could be made by including the proper band gap value  and temperature effect on $\mu$. But, now the question arises what could be the right band gap value of pure ZrNiSn? Because, the experimental values of  $S$ and $\rho$ as reported by Schmitt \textit{et al}\cite{snydertrue} at 300 K are $\sim$ -325 $\mu$VK$^{-1}$ and $\sim$1x10$^{-3}$  $\Omega$ m, respectively. And, the magnitude of $S$ in Ref. 14 is decreasing with increase in temperature and $\rho$ values are higher which indicate that band gap in ZrNiSn sample may be different compared to that of Shen \textit{et al.}\citep{shenexp} Also, for instance, in the experimentally prepared Heusler samples, there are chances of Ni deficiency leading to non-stoichiometric compound or mutual substitution between the Sn and Zr atoms.\cite{nidef} This defect or disorder may lead to reduction in the overall gap of the compound. In order to check the  effects of defect and disorder previously mentioned in ZrNiSn we performed the density of states calculations using Korring-Kohn-Rostoker (KKR) method under coherent potential approximation.\cite{kkr} The calculations showed the appearance of electronic states in the gap region leading to reduction of the effective gap. The effect of Sn-Zr substitutional disorder on the size of gap studied by {\"O}{\u{g}}{\"u}t \textit{et al.} also showed the reduction in effective gap.\cite{rabe} Therefore it appears that experimentally prepared samples in Ref. 15  is not pure (non-stoichiometric or disordered) and the band gap in pure ZrNiSn may be higher than the reported experimental value. The SCAN and mBJ calculations are predicting that pure ZrNiSn may have band gap of $\sim$0.54 eV. Therefore, it is interesting to study the thermoelectric properties of pure ZrNiSn and to see how they can be further enhanced. In this direction the thermoelectric properties of pure ZrNiSn are first systematically predicted. For the calculations, the electronic relaxation time $\tau_{e}$ are obtained by fitting the experimental $\sigma$ and  $\kappa_{e}$ data. Later, the possibility of achieving  higher $ZT$ and efficiency in ZrNiSn based compounds by doping are explored through power factor calculations.

\begin{figure}
\includegraphics[width=7cm, height=5cm]{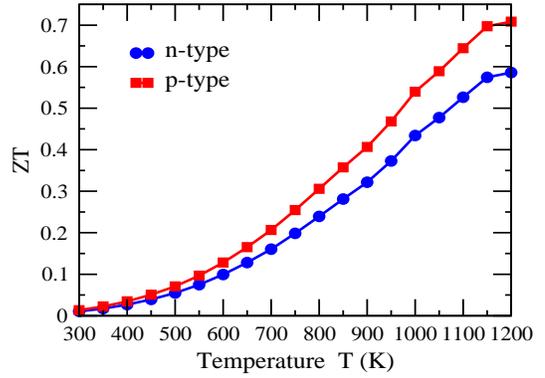} 
\caption{ The figure of merit $ZT$ for the doped ZrNiSn compounds as a function of temperature T. }
\end{figure}

 The obtained electronic transport properties of ZrNiSn with band gap of 0.54 eV are presented in Fig. 3. The Seebeck coefficient of pure ZrNiSn is shown in Fig. 3 (a). The value of $S$ is positive suggesting that contribution to thermoelectric voltage in ZrNiSn are mainly from holes even upto 1200 K. The maximum value of $S$ is $\sim$750 $\mu VK^{-1}$ at 300 K and starts to reduce with the increase in temperature. At 1200 K, the values of $S$ reaches $\sim$200 $\mu VK^{-1}$. The calculated  $\sigma$ and  $\kappa_{e}$ of ZrNiSn with temperature are shown in   Fig. 3 (b) and (c), respectively. Both the $\sigma$ and $\kappa_{e}$ show an increasing behaviour with temperature. Upto 600 K, the value and change in $\sigma$ (or $\kappa_{e}$) is lower but as temperature increases large increase in $\sigma$ (or $\kappa_{e}$) can be observed due to the excitation of more charge carriers across the band gap as in the case of a semiconductor. 

\begin{figure}
\includegraphics[width=7cm, height=5cm]{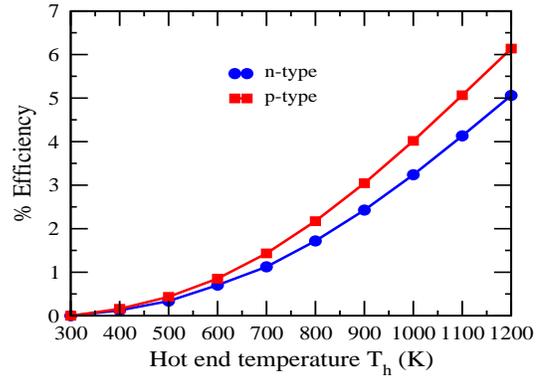} 
\caption{ The \% efficiency of n-type and p-type modules as a function of hot end temperature $T_{h}$. }
\end{figure}

\begin{table}[b]
\caption{\label{tab:table1}%
The effective mass ($m^{*}$) of charge carriers calculated in bands B1 to B4 for four k-path directions in the neighbourhood of $\Gamma$ and $X$ points.
}
\begin{ruledtabular}
\begin{tabular}{lcccc}
\textrm{}&
\textrm{$\Gamma$-$\Gamma$L}&
\textrm{$\Gamma$-$\Gamma$X}&
\textrm{X-X$\Gamma$}&
\textrm{X-XW}\\
\colrule
B1 & 0.28 & 0.40 & - & -\\
B2 & 2.42  & 1.00 & 1.51 & 0.21  \\
B3 & 2.42 & 1.00 & 1.51 & 2.50\\
B4 & - & - & 3.56 & 0.43\\
\end{tabular}
\end{ruledtabular}
\end{table}

Enhancing the power factor of a thermoelectric material by doping is one of the strategies to get higher figure of merit. Theoretically, one can predict the value of doping or proper carrier concentration that enhances power factor through the power factor versus chemical potential plot. In order to find out the optimal carrier concentration values which can yield higher power factor, we have plotted $S^{2}\sigma$/$\tau$ (PF) as a function of  $\mu$ in Fig. 3 (d). The Fig. 3(d) shows PF curves at temperatures from 300 to 1200 K. In the figure, the consecutive temperatures from 300 K to 1200 K  are to be read in the order of peak height. As can be seen from the figure there are two peaks corresponding to maximum PF at 1200 K. In the n-type region (positive $\mu$ levels) of doping the maximum PF peak is at $\mu$ level of $\sim$435 meV. The optimal electron doping corresponding to this $\mu$ level of $\sim$435 meV is $\sim$7.6 x 10$^{19}$ cm$^{-3}$ which gives the maximum power factor. At 1200 K, the value of PF at $\sim$435 meV is $\sim$100 x10$^{14}\mu WK^{-2}cm^{-1}s^{-1}$. Similarly, in the p-type doping region the maximum value of PF can be obtained at $\mu$ of $\sim$401 meV below the middle of the gap. For this value of $\mu$ representing hole doping the maximum PF value is $\sim$130 x10$^{14}\mu WK^{-2}cm^{-1}s^{-1}$ which is higher than PF obtainable from electron doping in this compound. The optimal value of hole concentration needed to attain this maximum PF in ZrNiSn related to $\mu$ level of $\sim$ -401 meV is $\sim$1.5 x 10$^{21}$ cm$^{-3}$.

Further, to evaluate the material for thermoelectric application we calculated figure of merit, $ZT$ for the doped ZrNiSn compounds with optimal electron and hole doping estimated before. The lattice part of thermal conductivity $\kappa_{ph}$ required to calculate $ZT$ is obtained by subtracting electronic contribution $\kappa_{e}$ (Fig.2 (d)) from the total thermal conductivity of ZrNiSn from Ref. 15. 
The $\kappa_{e}$ and $S^{2}\sigma$ values at the $\mu$ corresponding to the optimal carrier concentrations are obtained using the electronic relaxation time obtained earlier. The calculated $ZT$ for both the n-type and p-type doping corresponding to $\sim$7.6 x 10$^{19}$ cm$^{-3}$ and $\sim$1.5 x 10$^{21}$ cm$^{-3}$, respectively are shown in Fig. 4 for the temperature range of 300-1200 K. As can be seen from the figure, the $ZT$ value is increasing with temperature. This nature of $ZT$ suggests doped ZrNiSn compounds can be used for application in high temperature region. The figure of merit for p-type doping is higher compared to $ZT$ due to n-type doping. At 1200 K,  the value of $ZT$ are $\sim$0.6 and $\sim$0.7 for n-type and p-type compounds, respectively. 

The values of $ZT$ predicted are fairly high for the doped ZrNiSn compounds to be considered for high temperature thermoelectric applications. But, here it is important to note that in predicting the $ZT$, we have used the $\kappa_{ph}$ of the undoped ZrNiSn from Ref. 15 which is higher compared to its doped compounds. As can be seen in Ref. 15, doping the Zr site with heavy element Hf and Ni site with Pd one can systematically reduce the thermal conductivity and enhance $ZT$. So, if we find out $\kappa_{ph}$ for the doped sample from the experimental data and consider in $ZT$ calculation, then the predicted $ZT$ for p-type and n-type ZrNiSn compounds at 1200 K will be enhanced to $\sim$1.4 and $\sim$1.2, respectively. These value of $ZT$ obtained are higher compared to that of the maximum $ZT$ ($\sim$0.7) obtained in the work of Shen \textit{et al.}\cite{shenexp} Therefore, our work highlights the importance of considering doping for the pure ZrNiSn half-Heusler (band gap of 0.54 eV) to get further improvements in $ZT$.


Since, both the n-type and p-type doped ZrNiSn compounds have reasonably good value of $ZT$, it is desirable to consider the two materials for high temperature applications. The n-type and p-type materials can be used as n- and p-type legs in thermoelectric generators (TEGs). Therefore, the assessment of a material for TEG application can be done by calculating its efficiency. An approach of  segmentation to calculate the efficiency of thermoelectric materials was given in the work of Gaurav \textit{et al}.\cite{gaurav} In this method the module is divided into number of segments (or slices) depending on the hot and cold end temperatures ($T_{h}$ and $T_{c}$, respectively) and temperature difference $\Delta T$ across the each segment. The details of the methodology can be found in the Ref. 8. The efficiency of each module is calculated by keeping the $T_{c}$ fixed at 300 K  and varying the $T_{h}$ in steps of 100 K upto 1200 K. The efficiency thus calculated for both n-type and p-type legs are plotted in Fig. 5. As can be seen from the figure the \% efficiency of p-type module is higher than the n-type module. For the hot end temperature of 1200 K,  the efficiency of p-type module  is $\sim$6.1 \% and that of n-type module is $\sim$5.1 \%. It is to be noted that if the improved $ZT$ values estimated earlier are used, further enhancement in the efficiency can be obtained.  

The efficiency is slightly lesser than Bi$_{2}$Te$_{3}$ or PbTe\cite{gaurav}, but the ZrNiSn compounds have other advantages as half-Heusler materials. This compound is thermally stable, has high melting point\cite{jungmp}, easy to produce and due to the non-toxic nature of elements the ZrNiSn based compounds can be considered for high temperature thermoelectric applications.\cite{felser} For a proper design of a TEG, considering the thermal expansion of the material becomes an important aspect. Thus, in order to calculate the thermal expansion coefficient and to see how good the thermal expansion in ZrNiSn (which is an unharmonic effect) can be explained by the forces calculated through DFT with supercell-FDM approach and quasi-harmonic approximation, we have performed phonon calculations. The obtained phonon related properties and thermal expansion of ZrNiSn are discussed in the next section.

\subsection{\label{sec:level2}Phonon properties and thermal expansion}
The phonon dispersion of ZrNiSn compound is calculated along high symmetric directions which is shown in Fig.6. There are six optical and three acoustic phonon branches as can be seen in the dispersion plot for the ZrNiSn half-Heusler. The number of optical phonon modes per $k$-point are three lesser compared to that of the full Heuslers with formula unit of type X$_{2}$YZ. The optical and acoustic phonon branches are separated by a small gap of $\sim$1.5 meV. Around 30 meV there are three branches corresponding to optical phonons of higher energy. 

\begin{figure}
\includegraphics[width=8cm, height=5cm]{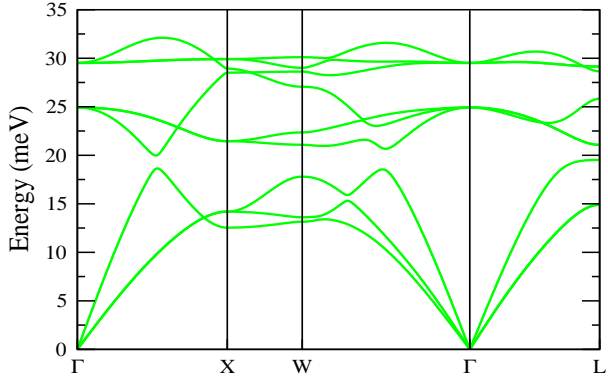} 
\caption{The phonon dispersion of ZrNiSn compound.}
\end{figure}

Further, we have calculated phonon total density of states (DOS) and  partial DOS to see the contribution of different atoms to acoustic and optical vibrational modes of the spectrum. The Fig. 7 shows the phonon total DOS per unit cell and phonon partial DOS per atom for ZrNiSn compound. The total DOS plot shows, in the region around $\sim$13, $\sim$22, $\sim$30  meV, three main peaks, respectively. As can be seen from the phonon partial DOS plot, the acoustic vibrational modes are predominantly from the heavier Sn atom in the formula unit. The Ni atom being lighter in mass contributes significantly to the higher energy optical phonons in the neighbourhood of $\sim$30 meV. 

\begin{figure}
\includegraphics[width=8cm, height=5cm]{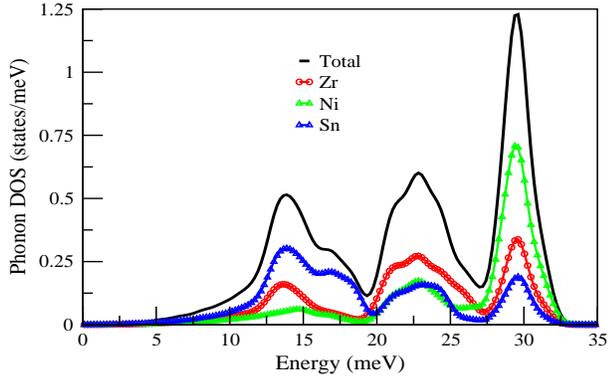} 
\caption{The phonon total DOS per unit cell and partial DOS per atom for ZrNiSn. }
\end{figure}

\begin{figure}
\includegraphics[width=7cm, height=5cm]{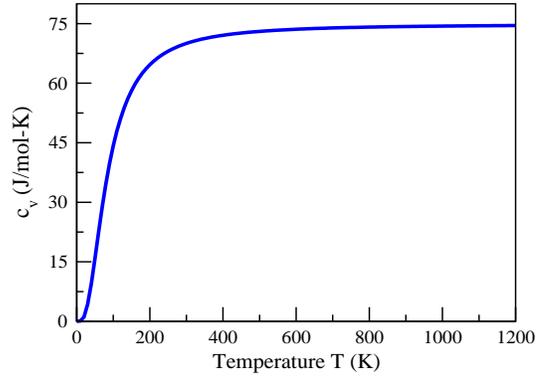} 
\caption{The constant volume specific heat c$_{v}$ as a function of temperature. }
\end{figure}

\begin{figure*}
\includegraphics[width=14cm, height=7cm]{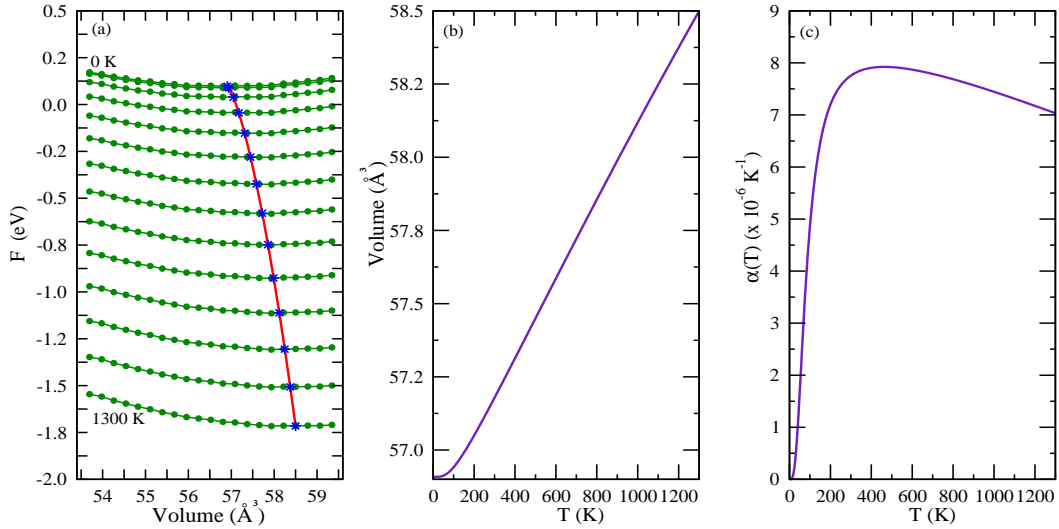} 
\caption{(a) Variation of total free energy $F$ with primitive cell volume. (b) Primitive cell volume variation with temperature T. (c) Linear thermal expansion coefficient $\alpha(T)$ with temperature T for ZrNiSn.  }
\end{figure*}
The Debye frequency $\omega_{D}$ which is a measure of maximum phonon frequency is $\sim$33 meV in case of ZrNiSn as computed from total DOS. Using the value of $\omega_{D}$, the calculated Debye temperature $\Theta_{D}$ for ZrNiSn is $\sim$382 K. The value of $\Theta_{D}$ indicates that all the vibrational modes in ZrNiSn are excited above 382 K. The value of $\Theta_{D}$ for ZrNiSn is relatively lower compared to the $\Theta_{D}$ of full Heusler compounds studied in our previous work.\cite{paper2,paper3} Rogl \textit{et al.} have reported $\Theta_{D}$ of 398 K for ZrNiSn compound from sound measurements\cite{debyetemp}. The value of $\Theta_{D}$ calculated in the present work is in quite good agreement with the experimental value. The variation of phonon part of specific heat at constant volume $c_{v}$ with temperature calculated under harmonic approximation is shown in Fig. 8. As can be seen from the plot, above 400 K the variation in $c_{v}$ is almost constant and is quite close to the Dulong-Petit value of $\sim$75 Jmol$^{-1}$K$^{-1}$.

For any material involving in applications where it is subjected to large temperature variations, understanding of its thermal expansion behaviour becomes important. In the design of TEGs where the materials used in modules experience temperature gradient, thermal expansion parameter needs to be taken into account. Therefore, a study of the thermal expansion in ZrNiSn is carried out in the present work. The linear thermal expansion coefficient $\alpha(T)$ of ZrNiSn compound is calculated under quasi-harmonic approximation.

  The calculated $\alpha(T)$ for ZrNiSn from temperature 0 to 1300 K is presented in Fig. 9 (c). The Fig. 9(a) shows, the variation of total free energy with change in volume of primitive unit cell, at a given temperature. Here, the total free energy at a given temperature $T$ and particular volume $V$ is defined as $F = [U_{el}(V)-U_{el}(V_{0})]+F_{ph}(T;V)$. The $F$ obtained is the sum of the Helmohltz free energy of the phonon part $F_{ph}(T;V)$ and relative ground state electronic total energy $U_{el}(V)-U_{el}(V_{0})$, where $V_{0}$ is equilibrium volume $V_{0}$ at 0 K. This $F$ vs. volume plot is useful in obtaining the equilibrium volumes at different temperatures. For each curve in Fig. 9 (a), the minimum $F$ point gives equilibrium volume at that temperature. In Fig. 9 (a) each such minimum energy point is connected through the solid (red) line from temperature 0 to 1300 K. Using the obtained equilibrium volume at each temperature from Fig. 9 (a), the variation of primitive cell volumes of ZrNiSn compound with increase in temperature is plotted in Fig. 9 (b). Above $\sim$100 K the volume is increasing monotonically with temperature. The volume of ZrNiSn at 1300 K is increased by an amount of $\sim$2.3\% from its room temperature volume. In order to compare with the experimental thermal expansion data in the work of Jung \textit{et al}\cite{jungmp}, we calculated average linear thermal expansion coefficient from the volume vs. temperature data in Fig. 9 (b). The average linear thermal expansion coefficient is calculated using the below formula:
\begin{equation}
\alpha_{ave}(T)= \frac{1}{a_{RT}} \frac{a(T)-a_{RT}}{T-T_{RT}},
\end{equation}  
here, $a_{RT}$ is lattice constant at room temperature $T_{RT}$ and $a(T)$ is lattice constant at temperature $T$. The value of $\alpha_{ave}(T)$ calculated is $\sim$7.8x10$^{-6}$ K$^{-1}$ in the temperature range 300-1000 K. Experimental value of $\alpha_{ave}(T)$  from dilatometer measurement (313-963 K) is 12.1x10$^{-6}$ K$^{-1}$ and from high temperature XRD measurement (323-673 K) is 11.0x10$^{-6}$ K$^{-1}$.\cite{jungmp} The value of $\alpha_{ave}(T)$ calculated under quasi harmonic approximation using the forces calculated from DFT with supercell-FDM approach is in quite good agreement with the experimentally reported values.

Using the primitive cell volumes at different temperature data in Fig. 9 (b) first, the volumetric thermal expansion coefficient is calculated. The volumetric thermal expansion coefficient is given by $\beta(T)= \frac{1}{V(T)} \frac{\partial V(T)}{\partial T}$. For the cubic structure, considering uniform expansion in three directions\cite{ashcroft} the linear thermal expansion coefficient $\alpha(T)$ can be obtained from $\beta(T)$ as $\alpha(T)= (1/3)\beta(T)$. Thus obtained $\alpha(T)$ in the 0 to 1300 K temperature range for ZrNiSn is presented in Fig. 9 (c). The value of $\alpha(T)$ for ZrNiSn in rising rapidly till $\sim$200 K and reaches maximum value of $\sim$7.9x10$^{-6}$ $K^{-1}$ around 450 K. After 450 K, thermal expansion rate in ZrNiSn is reducing upto 1300 K. The calculation of thermal expansion is a useful design parameter for TEGs. The calculated $\alpha(T)$ gives a measure of the extent to which change in length occurs in the thermoelectric material. Thus, one can take into account this parameter when designing a TEG module or a  hybrid TEG using segmentation of two materials\cite{gaurav2} to reduce the possible stress and cracking due to thermal expansion.

\section{Conclusions} 
In our work, electronic structure of ZrNiSn is calculated using SCAN XC functional. The effective mass values are calculated at the conduction and valence band extrema under parabolic approximation. The obtained indirect $E_{g}$ of $\sim$0.54 eV from SCAN and mBJ are in agreement with previous reports. The experimental $S$ values of a ZrNiSn sample is explained using DFT and transport calculations. Using the predicted $E_{g}$ of $\sim$0.54 eV the experimental $S$ is not explained. It is observed that in understanding the experimental $S$ of the sample, considering the $E_{g}$ of 0.18 eV and including temperature dependent shift in $\mu$ are important.  The need for considering $E_{g}$ of 0.18 eV to understand $S$ suggested that synthesized sample is not ordered and motivated to explore possible improvement in TE properties from ordered ZrNiSn. For the ZrNiSn compound with 0.54 eV band gap the thermoelectric properties are systematically predicted using the $\tau_{e}$ estimated from the experimental data and by considering temperature dependent shift in $\mu$. The possibility of enhancing the thermoelectric properties by considering doping for the ZrNiSn (band gap of 0.54 eV) is explored. The optimal carrier concentration required to achieve highest power factor for n-type and p-type ZrNiSn are found to be $\sim$7.6x10$^{19}$ cm$^{-3}$ and $\sim$1.5x10$^{21}$ cm$^{-3}$, respectively. The highest $ZT$ for the n-type and p-type compounds are $\sim$0.6 and $\sim$0.7 at 1200 K, respectively. By doping with heavy elements to reduce $\kappa$ the $ZT$ are predicted to be enhanced to $\sim$1.4 and $\sim$1.2 for p-type and n-type compounds, respectively at 1200 K. Therefore, it is observed that thermoelectric properties are much enhanced by considering doping in ZrNiSn with band gap 0.54 eV compared to that in band gap 0.18 eV. With reasonably high value of $ZT$, \% efficiency for n-type and p-type  materials are calculated for applying in the legs of TEGs. The obtained efficiency are $\sim$6.1 \% and $\sim$5.1 \% for p-type and n-type module, respectively. Further, phonon dispersion, total and partial density of states of ZrNiSn compound are studied under harmonic approximation. The value of $\Theta_{D}$ (382 K) calculated using $\omega_{D}$ obtained from phonon DOS is in good agreement with experimental value of 398 K. Under QHA, the thermal expansion behaviour in ZrNiSn is studied. The value of $\alpha_{ave}(T)$ of $\sim$7.8x10$^{-6}$ K$^{-1}$ calculated in our work is in quite good agreement with the experimentally reported values.

\section{Acknowledgements}
The authors thank Science and Engineering Research Board (SERB), Department of Science and Technology, Government of India for funding this work. This work is funded under the SERB project sanction order No. EMR/2016/001511.

\section{References}
\bibliography{ref}
\bibliographystyle{apsrev4-1}

\end{document}